# Deterministically Deterring Timing Attacks in Deterland

Weiyi Wu

Yale University
New Haven, CT, USA
weiyi.wu@yale.edu

Bryan Ford

Swiss Federal Institute of Technology (EPFL)
Lausanne, Switzerland
bryan.ford@epfl.ch

**Abstract**

The massive parallelism and resource sharing embodying today's cloud business model not only exacerbate the security challenge of timing channels, but also undermine the viability of defenses based on resource partitioning. We propose *hypervisor-enforced timing mitigation* to control timing channels in cloud environments. This approach closes "reference clocks" internal to the cloud by imposing a deterministic view of time on guest code, and uses timing mitigators to pace I/O and rate-limit potential information leakage to external observers. Our prototype hypervisor is the first system to mitigate timing-channel leakage across full-scale existing operating systems such as Linux and applications in arbitrary languages. Mitigation incurs a varying performance cost, depending on workload and tunable leakage-limiting parameters, but this cost may be justified for security-critical cloud applications and data.

## 1. Introduction

The cloud computing model, which shares computing resource among many customers, has become popular due to perceived advantages such as elasticity, scalability, and lower total cost of ownership. Clouds typically use virtualization to protect one customer's computation and data from another. Virtualization alone does not protect against timing attacks [40, 57], however, in which a malicious party learns sensitive information by analyzing the observable timing effects of a victim's computation.

Computers contain many shared resources that can be used for internal timing attacks by a co-resident attacker. L1 data caches [39], functional units [48], branch target caches [2], and instruction caches [1] have all been used to learn sensitive information. Protecting these resources individually through partitioning [18, 36, 55] is insufficient as attackers regularly identify new timing channels, such as L3 caches, memory, and I/O devices. Hardware features such as address space layout randomization attempt to make attacks more difficult, but often fail to do so adequately [28]. Even external attackers on a remote machine can exploit side-channels to learn a server's private key [12], and possibly any information not processed in constant time [10].

Other existing defenses against timing channels rely on language support and impose new programming models or limits to communication with the external world, making them unusable on unmodified existing application code. Deterministic containers rely on language support [4, 42, 45, 55], and Determinator requires applications to be ported to new microkernel APIs [6, 7]. Predictive mitigation offers important formal techniques to reason about and limit leakage [5, 45, 53, 54], but to our knowledge have not been implemented in a fashion compatible with existing languages and applications.

Today's cloud consumers expect to execute unmodified applications in a variety of existing languages, and to have unrestricted network access, precluding most existing techniques. Practical timing channel protection in cloud environments must support unmodified applications, unrestricted network access, and multi-tenancy.

As a first step towards this goal, we introduce Deterland, a hypervisor that builds on many of the techniques summarized above, but can run existing guest OS and application code in an environment guaranteeing strong rate-limits to timing channel leakage. Deterland uses system-enforced deterministic execution to eliminate internal timing channels exploitable by guest VMs, and applies mitigation to each VM's I/O to limit information leakage to external observers.

Deterland permits unmodified VMs and applications to run in parallel on the same machine and permits remote network interaction, while offering fine-grained control over the maximum information leakage via timing channels. Deterland presents guest VMs with a deterministic internal notion of time based on instructions executed instead of real time, and mitigates external leakage via techniques similar to predictive mitigation but adapted to the hypervisor's I/O model.

We have implemented Deterland by extending CertiKOS [24], a small experimental hypervisor. Unsurprisingly, we find that Deterland's general and backward-compatible timing channel defense comes with performance cost. However, cloud providers and their customers can tune Deterland's mitigation parameters for different tradeoffs between performance and leakage risk. Deterland's overhead can be fairly small (*e.g.*, less than 10%) in some configurations on compute-intensive or latency-tolerant workloads, while mitigation is more costly for interactive or latency-sensitive workloads. We expect these costs could be reduced with further hypervisor improvements, and could also benefit from hardware features such as precise instruction counting.

While timing channel mitigation via Deterland is probably not suitable for all cloud environments, our results suggest that it may offer promise as a general-purpose, backward-compatible approach to harden security-critical cloud computations whose sensitivity justifies the performance cost of mitigation.

In summary, this paper's primary contributions are 1) the first hypervisor offering general-purpose timing channel mitigation for unmodified guest operating systems and applications, 2) an evaluation of the overheads and leakage tradeoffs for various workloads, and 3) a tunable mitigation model enabling cloud providers and customers to control the tradeoffs between performance overheads and leakage risk.

## 2. Background

Timing channels [30, 51] require a victim-owned, attacker-accessible resource and a reference clock to determine access delays in order to infer sensitive information processed by the victim's computation. An attacker, co-resident to a victim's process, can exploit *internal timing channels* based upon shared resources [1–3, 32, 39, 48, 50, 57]. Alternatively, if a victim has a network, or otherwise externally, accessible resource, an attacker can exploit *external timing channels* by observing the timing relationship between inputs and outputs of the victim's resource [10, 12].

### 2.1 Internal Timing Channels

Figure 1 illustrates a typical *internal timing attack*, where an attacker attempts to learn a victim's secret key as the victim runs known code such as AES encryption. The attacker begins by selecting a shared resource, such as the L1 cache, and a time source, such as a high precision timer. In the case of an L1 cache, the attacker determines which cache blocks the victim accesses during known code sequences, *e.g.*, while performing key-dependent table lookups in AES. To obtain this information, the attacker first touches each of the L1 cache blocks to evict any data held by the victim, then yields the CPU to the victim. Upon resuming, the attacker then reads from the cache and measures the time for the reads to complete. Cache misses will take longer than cache hits, revealing the victim's memory accesses. This information has been used to learn AES keys [3, 50], ElGamal keys [57], and other sensitive cryptographic content [32].

Internal timing attacks require access to a high-resolution clock, common in existing computer architectures and made available to userspace by most operating systems. Even if the OS or hypervisor were to disable direct high-resolution timing sources such as this, hardware caches and interconnects [1, 2, 39, 48] as well as I/O devices [30, 51] are readily usable as reference clocks. Even a thread incrementing a shared-memory counter in a tight loop can create a reference clock [6, 51]. Resource partitioning can eliminate sharing of cache [55], memory [36], and I/O [18], but partitioning undermines the sharing and oversubscription advantages of the cloud model [6] and perpetuates an arms race as adversaries identify more resources usable for timing channel attacks.

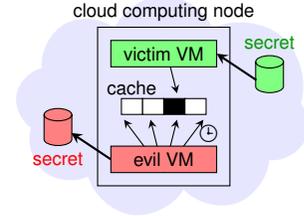

Figure 1: Internal timing channel. The attacker first flushes the cache, then the victim's VM performs memory operations whose access pattern depends on a secret. The attacker finally uses a high-precision clock to detect the victim's cache accesses, thereby deducing the secret.

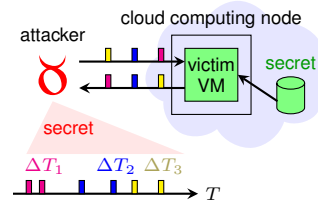

Figure 2: External timing channel. The victim's VM offers a SSL-secured Web service. The attacker sends a series of SSL requests and externally measures the victim's response delays to deduce the victim's secret.

Methods for addressing internal timing channel attacks include language-based techniques and system-enforced determinism. Language-based techniques [4, 42, 45, 54] eliminate reference clocks through time-oblivious operations that effectively execute for a fixed period of time regardless of input. Alternatively, system-enforced determinism [6, 7] eliminates reference clocks by isolating processes from sources of time both internal and external. Clocks accessible to untrusted code reveal only the number of instructions executed, and communication occurs at explicitly-defined points in time, preventing the modulation of a reference clock. While useful, these approaches impose new programming environments, breaking backwards-compatibility with existing code bases, and do not address external timing attacks.

### 2.2 External Timing Channels

Figure 2 illustrates a typical *external timing attack*, where an attacker attempts to learn a victim's RSA private key through network-based communication. The attacker initiates several SSL sessions with a victim, causing the victim to perform repeated RSA calculations using the same private key. If the victim's RSA arithmetic is not coded with sufficient care, the time the victim takes for session initiation statistically reveals information about the victim's private RSA key. By measuring response delays against a local reference clock, the attacker eventually deduces the victim's private key [12]. This style of attack generalizes to many cryptographic operations not implemented in constant time [10].

One way to address external timing channels is to delay all output from the system. Early proposals added random noise [23, 26], but offered no formal bounds on information leakage. More recently, predictive mitigation [5, 53] delays events according to a schedule predicted using only non-sensitive information. Mispredictions incur a doubling of future delays, however, which may result in unbounded performance penalties. Additive noise and predictive mitigation do not directly counter timing attacks in cloud environments, in which an attacker co-resident with the victim [56] can use fine-grained timers to exploit shared resources [40].

## 3. System Overview

Before delving into design details, this section presents a high-level overview of the Deterland architecture. We use the word *Deterland* to refer both to this architecture and more generically to a cloud environment using Deterland as its hypervisor. The name Deterland was inspired by *Neverland* [9], a mythical place where time stops, and as such must be inherently free of timing channels.

### 3.1 Mitigation Domains

A Deterland cloud, shown in Figure 3, may contain multiple *mitigation domains*. Virtual machines (VMs) in different mitigation domains receive different levels of protection against timing channels, or more precisely, different rate-limits on information leakage. As we will see in the evaluation (§6), mitigation inevitably comes with performance costs; a choice among several mitigation domains enables customers to balance performance against leakage risks.

Each mitigation domain has two properties for protecting against timing attacks: *mitigation interval* and *virtual CPU speed*. VMs within a given mitigation domain delay I/O that occur within a specified period until the end of that period constituting a *mitigation interval*. The virtual CPU speed is defined as the number of guest instructions executed within the guest VM per mitigation interval.

The cloud provider is responsible for creating mitigation domains for customers according to their requirements and contracts. Different mitigation domains may have different maximum information leakage rates, performance overheads and prices, but mitigation parameters are fixed within a mitigation domain. When a customer creates a VM instance, she chooses a mitigation domain suitable for a particular application. VM instances in different mitigation domains may be created interchangeably, and a VM instance can be migrated from one mitigation domain to another.

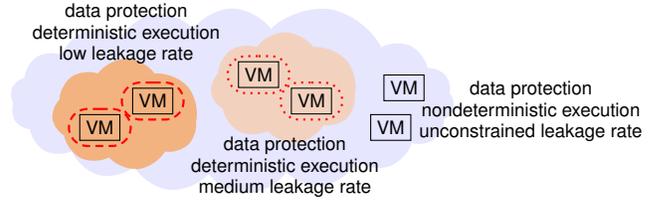

Figure 3: Mitigation domains in Deterland. Each color represents a different mitigation domain.

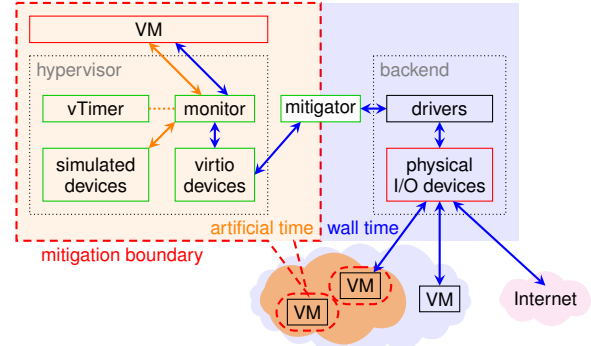

Figure 4: Deterland architecture overview. Orange components – including untrusted guest VMs – lie within the mitigation boundary, and hence have no direct access to real time. Blue components connect these orange components to the outside world, but all interactions between them pass through the mitigator.

### 3.2 Architecture Overview

Figure 4 illustrates Deterland's architecture, which is broadly similar to conventional hypervisors. However, Deterland enforces deterministic execution and predictive mitigation to control internal and external timing channels.

As shown in Figure 4, each VM instance running on Deterland has its own mitigation boundary, which delimits a specific mitigation domain. Code running on a guest VM within this boundary can observe only artificial time, based on the number of guest instructions executed since the last mitigation interval. Thus, the only internal timing source the VM can access is an artificial time generator (vTimer in Figure 4). The other components inside the boundary are also constrained to behave deterministically, so as to eliminate all internal timing channels.

The only nondeterminism that can affect computations in a mitigation domain are introduced via explicit I/O crossing the boundary, through the non-mitigated blue area in Figure 4. An external observer can measure the times at which outputs emerge from the cloud computation, yielding external timing channels. Deterland uses mitigation to limit the maximum rate of information leakage across this I/O boundary, however. An internal observer can receive time-stamped messages from an external clock source, of course, but mitigation limits the rate and granularity with which external time sources are observable by the VM.

Unlike many common hypervisors, Deterland does not allow its VMs to access hardware directly, but fully virtualizes all essential devices and isolates the VMs from the physical hardware. This device virtualization is necessary to prevent untrusted guest VMs from using nondeterministic device hardware to form high-precision internal timing sources, which would compromise mitigation security.

Deterland allows a guest VM to communicate with any remote endpoint, subject to other security policies the cloud provider might impose, but all external communications are mitigated – currently including communication between different cloud hosts within the same mitigation domain. An area for future improvement would be to support deterministic intra-cloud communication protocols that need not be mitigated, but we leave such optimizations to future work.

### 3.3 Security Assumptions

Deterland's goal is to protect against both internal and external timing attacks in an environment that can execute existing, unmodified operating system and application code. Specifically, Deterland offers timing channel mitigation against both co-resident and remote adversaries. An adversary is assumed to have full, unrestricted access to his guest VM's resources including all guest-accessible hardware and instructions. The adversary can run arbitrary code that can flush cache lines, attempt to measure memory access times, set and receive (virtual) timer interrupts, and send and receive network packets both inside and outside Deterland. We impose no special operational limits on the attacker; he may repeat these activities at will.

We assume an adversary can contrive to co-locate a process with a victim, and may allocate multiple virtual machines on the same machine, although Deterland applies mitigation to these VMs as a unit. The underlying hardware prevents direct access to the users' confidential data via virtualization and access controls. We assume an attacker cannot escalate privileges and compromise the cloud or a target's resource. An attacker can learn any public information about a target including virtual machine or process placement and published or declassified data.

Deterland assumes that clients employ secure communication primitives, such as HTTPS and TLS. An external attacker may have access to any I/O flows into and out of a hardware resource executing a target's process and can compare these accesses against a high-precision clock.

## 4. Deterland Design

The layout of Deterland components in a single cloud computing node is shown in Figure 5. The hypervisor runs virtual machines (VMs) inside a mitigation boundary, and isolates the VMs from real hardware. For implementation simplic-ity the mitigator and backend drivers currently run on separate CPU cores outside the mitigation boundary, although this current restriction is not essential to the architecture. The mitigator communicates with the backend drivers, and the backend drivers in turn access the underlying hardware. This section details the design of these three key modules: hypervisor (§4.1), backend (§4.2) and mitigator (§4.3).

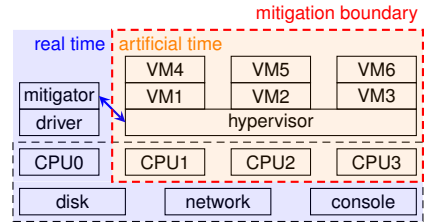

Figure 5: Layout of Deterland components

### 4.1 Hypervisor

The hypervisor consists of four main components: the VM monitor, the artificial time generator (vTimer), simulated non-I/O devices, and virtio devices, as shown in Figure 4. All four parts are in the Trusted Computing Base (TCB) because they can access VM state, which must not be visible to external observers. The hypervisor enforces deterministic execution and contributes to the mitigation process.

#### 4.1.1 VM Monitor (VMM)

The VM monitor in Deterland has functionality similar to a conventional VMM. In contrast with conventional VMMs designed to maximize computing power and minimize overhead, Deterland's VMM must enforce strict determinism within each guest VM despite the performance costs of doing so. This currently means that Deterland's VMM must virtualize all privileged instructions, I/O operations and interrupts, and avoid using hardware-assisted virtualization features that could compromise determinism. Common hardware virtualization features could in principle be enhanced to enforce deterministic execution, thereby potentially improving Deterland's performance considerably.

The VMM intercepts instructions that access timers, statistical counters, and other nondeterministic CPU state. Timing-related instructions are emulated by the vTimer, and statistical counters are currently inaccessible to guest VMs at all. The VMM in this way systematically prevents the guest from accessing CPU features that could be used as high-precision timing sources to obtain information about the execution timings of other guest VMs.

The VMM also tracks the precise number of executed instructions for the vTimer, and pauses the VM after a deterministic instruction count as dictated by the vTimer. Although some historical architectures such as PA-RISC [29] directly support exact instruction counting, on the x86 ar-chitecture the VMM must use well-known but more costly techniques to count instructions precisely in software [19].

Deterland does not support I/O passthrough, as the only physical hardware inside the deterministic boundary are CPU and memory, excluding timing-related features. All devices the VM can access are deterministically simulated. The VMM intercepts and forwards every I/O instruction to corresponding virtual device, either a simulated non-I/O device or a virtio device.

The VMM intercepts and handles all raw hardware interrupts, since hardware interrupts could be utilized as timing sources. The VMM therefore either disables physical hardware interrupts for the CPU core it runs on, or reroutes them to the CPU core the mitigators run on. The only interrupts the VMM injects into guest VMs are virtual interrupts generated deterministically by simulated devices.

### 4.1.2 vTimer

The artificial time generator, or vTimer, generates the only fine-grained notion of time observable to guest VMs, and schedules mitigated I/O and interrupts. Almost all other components refer to the vTimer's artificial notion of time.

The vTimer provides three APIs for other components: 1) get the current artificial time; 2) set an alarm at some future artificial time for an event, and 3) get the artificial time for the nearest event. The first API is used widely in other components as it is the only timing source they can access. The second API is used by simulated devices that take periodic actions. For example, the simulated timer device generates timer interrupts periodically; the virtio devices communicates periodically with the mitigator, at the end of each computation segment. The vTimer will notify the device at the artificial time so that it can take the corresponding action and reset the alarm for the next iteration. The third API is called by the VMM before each VM entry to pause the VM after the appropriate number of instructions are executed, or to manually advance the artificial time in some cases.

The vTimer updates the artificial time after each VM exit. The VMM notifies the vTimer with the number of instructions the VM executed since the last VM entries. The VMM also asks the vTimer to manually increase the artificial time when it manually advances the program counter in the case it intercepts an instruction and then "executes" it by hand. Some virtual devices may ask the vTimer to manually advance the artificial time in order to simulate I/O operations that normally take some time to complete. When the VM enters a halt state, the VMM asks the vTimer to advance the artificial time to the point for the next vTimer event.

### 4.1.3 Simulated Devices

The hypervisor simulates essential device hardware for guest VMs. The simulated devices behave deterministically, do not access the underlying hardware, and use the vTimer as their only timing source. Only the most critical devices are simulated, such as bus controller, interrupt controller and timer device, and Deterland implements a basic version of each device to minimize hypervisor complexity.

### 4.1.4 virtio Devices

All external-world I/O in Deterland uses the virtio interface [41]. Virtio is the widely-adopted, *de facto* standard for I/O virtualization in current hypervisors. Its request-based ring structure has a simple design, and its asynchronous nature enables I/O buffering and operation bundling across Deterland's mitigation boundary.

When a guest VM needs to perform I/O, it first creates descriptors and data buffers in guest memory, then notifies the virtio device. Conventional hypervisors extract requests from guest memory and process them right away. Deterland instead extracts the requests immediately but then buffers them in a mitigation queue. At the end of the current mitigation interval, the vTimer notifies the virtio device to send all the queued requests to the mitigator as a bundle, and receives a response bundle from the mitigator. After receiving this response bundle, the virtio device unpacks virtio responses and copies them into guest memory. When the virtio device finishes processing I/O responses, it sets a new alarm for the next mitigation interval.

The data transferred between the virtio devices and the mitigator are not the scattered, raw data buffers as in the virtio specification. For output requests, the virtio device allocates a shared data buffer in hypervisor memory, copies data from guest memory to the buffer, and sets up the request with the address of the buffer. This data buffer is shared between the hypervisor, the mitigator and the backend driver, but is inaccessible to the guest VM. For input requests, the virtio device simply sets up the request without an attached buffer. The data buffer is then allocated and attached to the corresponding response by the backend driver. For both types of requests, data buffers are attached to the responses so that the virtio device can safely deallocate the buffers after copying data into guest memory.

## 4.2 Backend

The backend performs actual I/O operations. After receiving an I/O bundle from the mitigator, the backend driver unpacks the bundle and performs the I/O requests it contains. Once an output request finishes, or an input arrives, the driver sets up the response and sends it back to the mitigator.

The backend driver does not bundle responses, because it does not know when a mitigation interval starts and ends. Instead the backend sends the mitigator individual responses and leaves the task of buffering and bundling responses to the mitigator.

## 4.3 Mitigator

The mitigator implements the core of the mitigation process, which adapts previously-proposed mitigation techniques to the hypervisor environment.

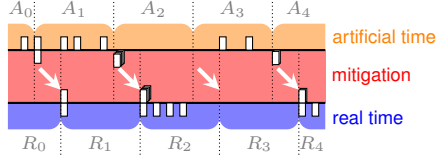

Figure 6: Mitigations for outputs: requests are bundled at the end of each artificial-time period, and become available for the device at the beginning of the next real-time period.

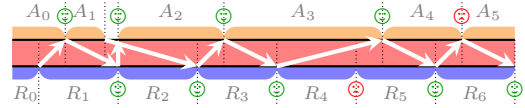

Figure 7: Information leakage: when no request bundle is available at the beginning of a real-time period, external observers gain one bit of information; when responses bundle contains responses from multiple real-time periods, the VM gains the same information. Red frownies represent potential one-bit information leakage events.

Deterland supports different mitigation intervals for different I/O queues of virtio devices for the same VM instance. In practice, however, we expect cloud providers to use the same mitigation interval for all virtio devices to rate-limit information leakage consistently within a mitigation domain. When a VM instance is created, the hypervisor synchronizes all mitigator instances for the virtio devices.

The control loop of a virtio device's mitigator is straightforward: 1) receive a request bundle from the virtio device; 2) collect all pending responses until the next time slot; 3) forward the request bundle to the backend driver, and 4) bundle all responses and send the bundle to the virtio device. The communication primitive the mitigator uses between the virtio device and itself is a synchronous exchange operation, while the primitive it uses between the backend driver and itself is an asynchronous send and receive.

As discussed in §4.1.4, the actual data buffer is shared among the mitigator, the virtio device and the local backend driver. Therefore the actual objects being mitigated are the I/O metadata, *i.e.*, virtio request bundles and response bundles and permissions for accessing the actual data buffer. For some hardware, the mitigator can be embedded in the local backend driver to further reduce communication overhead. In this case, the backend driver becomes a part of TCB.

### 4.3.1 Mitigation in Deterland

Deterland does not limit the number of I/O operations in a real-time period or an artificial-time period, unlike earlier mitigation designs [5]. By enforcing deterministic execution, the content and ordering of I/O operations in an artificial-time period become pure deterministic functions of the VM state and the contents of response bundles virtio devices receive at the beginning of the artificial-time period. We can therefore use the request bundle in place of a single request as the atomic unit of output mitigation, as shown in Figure 6.

Deterland mitigates both input and output. The mitigation for input is symmetric with that of output, and uses a response bundle as the unit of mitigation. Figure 7 shows the data flow of request and response bundles as well as information leakage. External observers can obtain at most one bit of leaked timing information in each mitigation interval. The VM may gain several bits of information in an artificial-time period, but no more than what external observers could have learned already. For example, in Figure 7, an external observer may learn one bit of information from the fact that "there is no output in $R_5$ because $A_3$ missed the deadline." Correspondingly, the VM learns that "the input at the beginning of $A_5$ comes from $R_4$ and $R_5$." These two "leaks" reveal exactly the same information, however, as they both reflect the same root cause – $A_3$ missed the deadline.

In practice, mitigation introduces 2–3 mitigation intervals of latency to I/O operations – the most important performance cost of mitigation. For example, in Figure 7, if the VM issues a disk I/O operation in $A_1$, the operation will be bundled at the end of $A_1$ and actually performed in $R_3$. The result of the operation will then be bundled at the end of $R_3$ and become available to the guest VM in $A_4$. The VM thus sees a 2–3 mitigation interval latency. If the VM misses deadlines during this roundtrip, the latency may be higher. Symmetrically, if an external request arrives in $R_1$, it can be processed as early as in $A_2$ and the corresponding response will be published as early as in $R_4$.

Multiple I/O devices for the same VM instance could leak more than one bit information per mitigation interval if incorrectly configured. This leakage is avoided if the cloud provider sets one mitigation interval for all VM instances in the same mitigation domain, and Deterland applies the value to all I/O queues for each VM instance. Deterland also keeps the real-time periods across I/O queues for the same VM instance synchronized. The artificial-time periods are naturally synchronized as they use the same notion of artificial time.

### 4.3.2 Mitigation Configuration

The maximum information leakage rate is directly related to the mitigation interval. However, the upper bound is likely to be hard to reach, especially when the computing node has low utilization. Actual information leakage is heavily impacted by the virtual CPU speed and other factors.

Figure 8 illustrates information leakage for different utilizations under different vCPU speed settings. In Figure 8a, the vCPU speed is significantly lower than the host CPU speed, which means that each segment runs only a small number of instructions and the host CPU can always complete them in a single real-time slot. Provided the VM always finishes the segment within one time slot, it effectively behaves like a hard real-time system and leaks no timing infor-

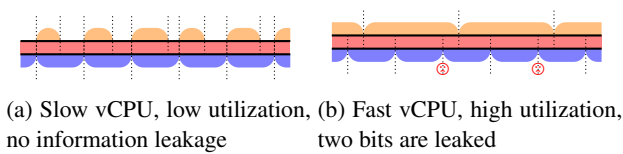

(a) Slow vCPU, low utilization, no information leakage

(b) Fast vCPU, high utilization, two bits are leaked

Figure 8: Information leakage and utilization

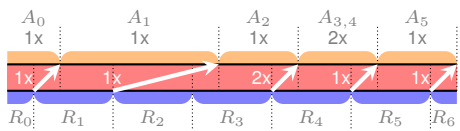

Figure 9: Time synchronization using piggybacked value from the mitigator

mation at all, at a likely cost of substantially under-utilizing the CPU. In contrast, the vCPU speed in Figure 8a is close to or higher than the host CPU speed. In this case the VM may leak one bit of information per time slot – namely whether the segment completed execution in this time slot or in some later slot – but utilizes the CPU more efficiently.

Calculating the actual amount of information an execution trace leaks using Shannon entropy [43] is possible in principle but difficult. However, we know that an under-utilized computing node is unlikely to miss deadlines and hence leak information. An over-utilized computing node, on which the VM has $50\%$ chance of finishing a segment in time and $50\%$ chance of missing the deadline, might be expected to leak information at the maximum rate.

#### 4.3.3 Time Synchronization

If one segment takes longer than expected, the artificial time the VM perceives will permanently become one mitigation interval behind real time, as $A_5$ and $R_6$ in Figure 9. Deterland automatically adjusts the guest VM's perception of time so that artificial and real time remain roughly synchronized, provided all I/O queues have the same mitigation interval. This requirement is normally satisfied as long as the cloud provider applies the same mitigation interval to all devices. This time synchronization leaks no additional information.

Just before exchanging buffers, the mitigator calculates the number of elapsed real-time periods since the last exchange, and piggybacks this number in the response bundle, as shown in Figure 9. The virtio devices are expected to receive the same number at the same time since the mitigators are synchronized, and the mitigation intervals for all devices are the same. After receiving this number, the virtio device notifies the vTimer to update the speed ratio accordingly.

The speed ratio controls how fast the vTimer advances artificial time. If the speed ratio of the vTimer is set to $n$, it advances the artificial time $n$ times faster than normal, which is equivalent to temporarily reducing the virtual CPU speed to $1/n$-th of its normal value. The number of instructions in a segment remains unchanged, however. Therefore the virtio devices need to set up the alarm $n$ times as far ahead as they normally would. This speed ratio compensates the missed real-time periods and quickly recovers artificial time to keep it synchronized with real time.

The only information crossing the mitigation boundary is the number of elapsed real-time slots, and this information is already known to the external observer and the VM, so the piggybacked value does not leak additional information. The maximum leakage rate remains constant since the actual length of segments and time slots remains unchanged.

### 4.4 Limitations

Deterland currently supports only a single virtual CPU per guest VM, though it can run many single-core guests on a multicore machine. Enforcing deterministic execution on multiprocessor guests is possible but complex and costly [20]. More efficient user-space libraries for deterministic parallelism [16] still introduce overhead and cannot prevent malicious guests from violating determinism to gain access to high-precision time sources. Determinator's microkernel-enforced deterministic parallel model [7] would satisfy Deterland's security requirements but is not compatible with existing applications or guest operating systems.

This limitation may not be a serious problem for many "scale-out" cloud applications, however, which are typically designed to divide work just as readily across processes or whole VMs as among shared-memory threads. Especially in large-scale parallel processing paradigms such as MapReduce, there may not be much fundamental efficiency difference between running a workload on $N$ $M$-core (virtual) machines versus running the same workload on $N \times M$ single-core (virtual) machines. Deterland can run these $N \times M$ single-core VMs organized as $M$ guests on each of $N$ physical $M$-core machines.

Another limitation of Deterland is that it mitigates all inter-VM communication, even among cloud hosts within the same mitigation domain, which is in principle unnecessary to ensure security. This limitation might be solvable through deterministic intra-cloud communication primitives or efficient deterministic-time synchronization techniques, but we leave such improvements to future work.

Deterland currently provides only coarse-grained timing information control, across entire guest VMs. An IFC-based timing information control [45] might in principle offer finer-granularity control and reduce mitigation overhead.

Finally, Deterland does not support direct hardware access from the VM even if the hardware itself supports virtualization features, such as Intel SR-IOV. Future hypervisor optimizations and hardware improvements could eventually enable Deterland to make use of hardware virtualization and reduce its performance costs.

## 5. Implementation

Deterland builds upon CertiKOS [24], a certified hypervisor. Deterland adds mitigation, an artificial timer, timing-related virtual devices such as vPIT and vRTC, and virtio devices. Deterland also virtualizes the network device via a network driver running in the hypervisor. CertiKOS gives guests direct access to network devices via IOMMU, but Deterland cannot do this without compromising determinism. Deterland also optimizes IPC performance for communication patterns resulting from I/O mitigation. Deterland's code additions to CertiKOS are not yet certified.

Deterland inherits several drawbacks from CertiKOS. CertiKOS currently only supports a single VM instance per CPU core. Also the CertiKOS disk driver utilizes a relatively slow AHCI mode.

Deterland uses Intel VT-x instructions for virtualization, but does not make use of Intel VT-d for direct I/O in order to isolate the VM from physical device hardware.

### 5.1 Precise Versus Coarse Execution

For experimentation purposes, the Deterland hypervisor can be configured to enforce either *precise* or *coarse* deterministic execution. The coarse mode reflects the currently-untrue assumption that the hardware is capable of interrupting execution after a precise number of instructions. Current x86 hardware unfortunately does not have this capability, but other architectures (notably PA-RISC) have offered it [29], and it could in principle be added to x86 processors. Coarse mode experiments may thus offer some insight into Deterland's potential performance on enhanced hardware.

In the precise execution mode, the Deterland hypervisor uses well-known but costly techniques to emulate exact instruction counting in software [19]. Deterland uses Intel's architectural performance monitoring counters to count the executed instructions in guest mode, and to pause the VM after a certain number of executed instructions. In practice the counter will nondeterministically overshoot, since the Intel CPU uses the local APIC to generate performance counter overflow interrupts instead of generating precise exceptions internally. The propagation latency is around 80 instructions in average, and the maximum latency we observed is around 250. Deterland therefore pauses the VM 250 instructions early, then uses Monitor TF to single-step the VM to the target instruction count.

Since the performance counters are highly nondeterministic, Deterland hides them from guest VMs by counterfeiting CPUID results and intercepting MSR accesses.

### 5.2 Artificial Time

The vTimer does not directly calculate artificial time, but only calculates instruction counts. The "time" the VM perceives is generated by the vRTC and vPIT, which use the virtual CPU frequency to calculate the number of instructions to execute between guest VM interrupts.

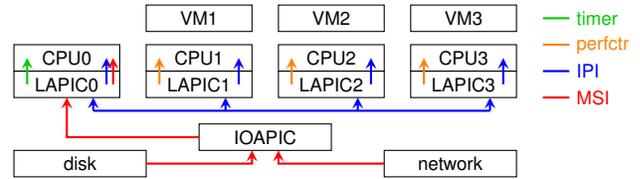

Figure 10: Interrupt routing

Deterland intercepts the `rdtsc` instruction and accesses to the MSR alias of the TSC. The vTimer then uses the number of executed instructions as the value of TSC and simulates a vCPU that executes exactly one instruction per cycle, to support OS and application code that relies on the TSC without compromising determinism.

### 5.3 Interrupts

Deterland disables the LAPIC timer, enables performance counter overflow interrupts on the CPU cores the VM runs on, and configures the IOAPIC to reroute all external interrupts to the core the mitigator runs on. The complete interrupt routing scheme is shown in Figure 10. Guest VMs are thus preempted by performance counter overflows, instead of by the true timer. Deterland uses IPIs for IPC since each guest VM runs under the control of a CertiKOS process.

Virtual interrupts are queued to the vPIC first, then injected into the VM as soon as IF in the guest's `eflags` is set. Virtualization of MSI is not yet implemented, and the interrupt numbers for devices are currently hardcoded.

### 5.4 I/O Virtualization

For simplicity, Deterland currently uses port I/O instead of memory-mapped I/O. MMIO and port I/O each cost one VM exit because Deterland intercepts all I/O operations. MMIO requires the VMM to decipher the guest page tables and instructions to get the actual I/O operation. Port I/O is much simpler, however, and the port address space is sufficient for all simulated and virtio devices.

## 6. Evaluation

This section evaluates the practicality of VM-wide timing mitigation, focusing on measuring the performance overhead of Deterland across different configurations. Experiments are run on a commodity PC system consisting of a 4-core Intel Core i5-4705S CPU running at 3.2GHz with 16GB of 1600MHz DDR3 RAM, a 128GB SSD and a gigabit Ethernet card (Realtek 8111G). This system design corresponds roughly to a low-cost cloud hardware configuration.

We compare execution on Deterland against QEMU/KVM 2.1.0 [31] running on Ubuntu 14.10. For a fair comparison, we enable VT-x and disable VT-d in the BIOS, so that KVM uses the same CPU features for virtualization. The guest OS is Ubuntu 14.04.2 LTS server version.

As Deterland, by design, does not expose real time to the VM, we measure performance overhead using vmcalls that report hypervisor statistical data to the VM via x86 registers. These calls are included strictly for experimentation and are not part of the production hypervisor interface.

We report performance overhead versus KVM along with an upper-bound estimate of information leakage rate. This estimated leakage is a loose bound reflecting total possible information flow across the mitigation boundary, including noise and known public information. It seems unlikely that practical data-exfiltration attacks could come anywhere close to this maximum leakage rate, except perhaps with the deliberate help of conspiring "trojan" code in the victim itself.

### 6.1 Cloud Configurations

Deterland depends on three main tunable parameters: mitigation interval, vCPU speed and precise/coarse mode. The vCPU speed may heavily impact the performance of CPU-bounded applications, as it limits the number of instructions executed in a time slot. The mitigation interval will likely impact the performance of I/O-bounded applications, as it adds delays for mitigation purposes. Precise mode introduces more overhead than coarse mode due to the need to perform precise instruction-counting in software.

We examine different settings and report both performance overhead and maximum information leakage. Information leakage rate does not strictly reflect Shannon entropy, but is approximated by the number of missed deadlines, assuming the VM normally finishes every segment in time. VM entry and exit overhead is categorized as VMX, and is measured by the number of VM exits. The time for of one VM exits is measured by single-stepping a piece of code. We use $1.0445$ microsecond per VM exit for all the benchmarks.

### 6.2 Micro-Benchmarks

The micro-benchmarks focus on three major computing resources a cloud provides: CPU (§6.2.1), network I/O (§6.2.2) and disk I/O (§6.2.3).

#### 6.2.1 CPU

The CPU benchmark uses the built-in CPU test in sysbench [33] to check 20000 primes. We tested different settings of vCPU speed and mitigation intervals. This experiment aims to reveal Deterland's overhead over KVM.

Figure 11a and Figure 12a shows performance results and maximum leakage. Figure 13a breaks down execution time. Performance slowdown is almost linear in vCPU frequency. Performance saturates near 6.4GHz because the CPU is capable of executing two simple instructions in one cycle, or 6.4 billion instructions per second.

Clearly the precise mode introduces extra overhead due to single-stepping. The overhead for single-stepping alone is 30% for 1ms mitigation interval, including the time spent in the hypervisor and during VM entries and VM exits. The same overhead is 5% in total for 100ms interval.

Almost all remaining overhead comes from mitigation, because the CPU-bounded application mostly executes instructions that do not cause VM exits. A cloud provider might in practice schedule another VM instance during the mitigation period of this VM. Thus the actual overhead to the cloud provider is small in coarse mode, or in precise mode with a long mitigation interval.

The higher red line in Figure 12a is the theoretical upper bound of information leakage rate for 1ms mitigation interval, while the lower red line is for 100ms mitigation interval. The estimated information leakage rate is significantly lower than the theoretical upper bound, and much of this estimated leakage is in turn likely to be noise.

#### 6.2.2 Network I/O

Network overheads for TCP and UDP were obtained via iperf [46], with default window sizes (47.3K for TCP and 160K for UDP). All tests run on a 3.2GHz vCPU.

Figure 11b and Figure 12b shows performance and maximum leakage. Figure 13b breaks down execution time. As an I/O bounded application, iperf does not execute the maximum number of instructions per time slot as in the CPU benchmark, because the kernel executes `hlt` when all processes are blocked by I/O.

The VM is under-utilized in most settings, but may still leak information. This leakage is caused by the hypervisor, as the hypervisor still needs to move data between guest memory and the shared data buffer. The mitigator may not receive bundled requests in time because the hypervisor is busy copying data. The estimated information leakage rate is much less than the theoretical maximum for TCP because traffic is low, while the leakage rate for UDP is higher due to higher-volume traffic. However, this estimated leakage mostly reflects public information about traffic volume, and is unlikely to contain much truly sensitive information.

UDP bandwidth changes over mitigation intervals as expected. When the mitigation interval is too short, the hypervisor spends significant time on VM entries and exits, instead of sending packets. In contrast, when the mitigation interval is too long, the VM may not have enough virtio descriptors to send packets, as most of them are buffered in the hypervisor or the mitigator.

TCP bandwidth exhibits more erratic behavior, because the guest's congestion control algorithms are severely impacted by the effects of mitigation. The reported result is measured with CUBIC; other algorithms aside from BIC are basically unusable. These results are understandable since congestion control is highly timing-sensitive. A future optimization might therefore be to move the TCP stack from the guest VM into the hypervisor and give the guest a more latency-tolerant socket-proxy interface.

#### 6.2.3 Disk I/O

*Raw disk*  The disk benchmark uses fio [8] for both synchronous disk I/O and Linux's native asynchronous disk I/O

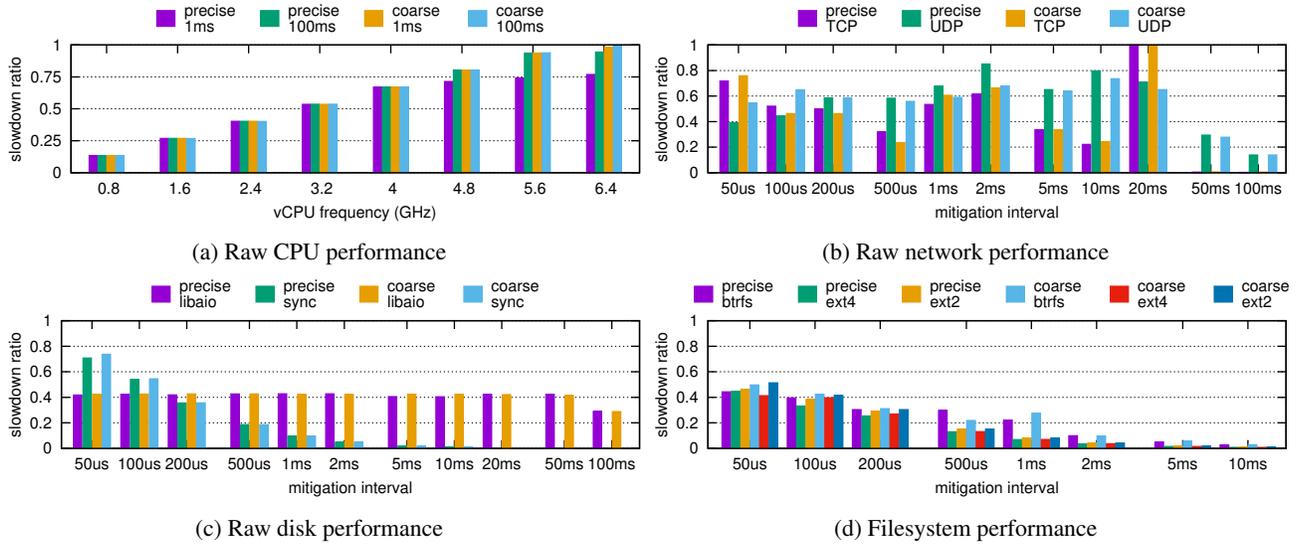

Figure 11: Performance overhead for micro benchmarks

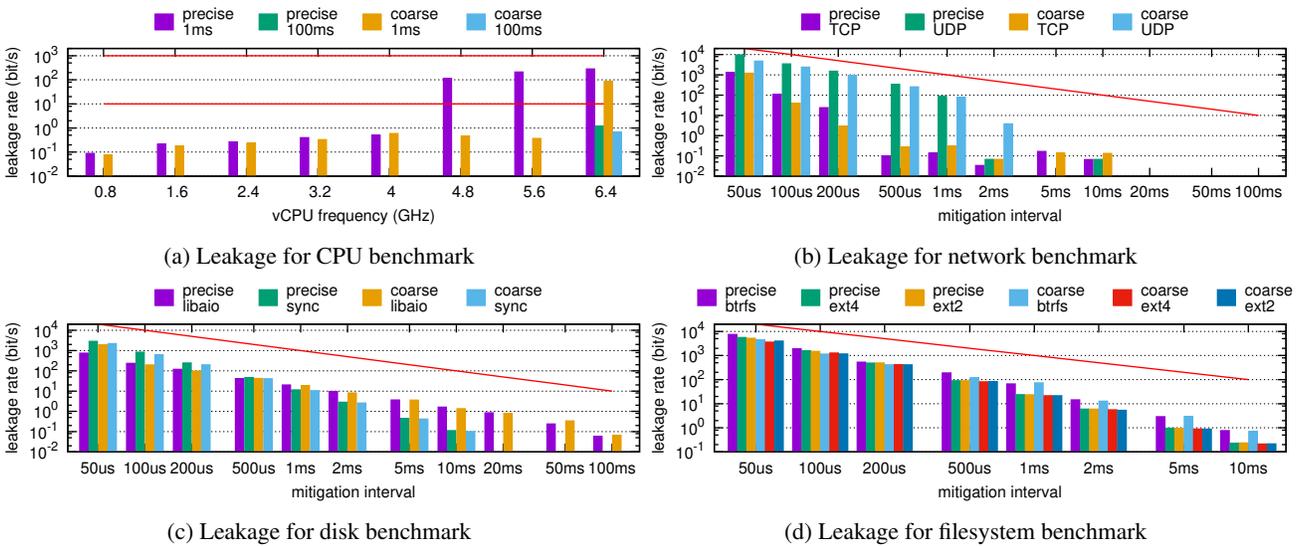

Figure 12: Estimated information leakage rate for micro benchmarks

(libaio). Synchronous disk I/O operations are expected to exhibit latencies up to three times as long as the mitigation interval, as discussed in §4.3.1. The total data size is 16GB for the libaio test, and 1GB for the synchronous test.

Figure 11c and Figure 12c shows performance results and maximum leakage. Figure 13c breaks down execution time.

The estimated leakage rate is a order of magnitude lower than the theoretical upper bound. As with the network benchmark, the estimated leakage rate is largely caused by data copying in the hypervisor, and is unlikely to contain much truly sensitive information.

The asynchronous disk I/O basically keeps the maximum rate for all mitigation intervals, except in the 100ms interval case due to a lack of virtio descriptors. Deterland is 50% slower than KVM because the underlying AHCI driver is not well-optimized. The throughput is still more than a spinning magnetic disk offers, however.

Synchronous disk I/O, on the other hand, shows a reciprocal rate as expected. With a 1ms mitigation interval, the throughput is 10 times slower than that of KVM, and the VM nearly stalls when the mitigation interval is more than 10ms. Fortunately, many disk I/O bounded applications, such as databases, use asynchronous I/O and data buffer pools to tolerate disk latency. Modern filesystems also have features like buffering and read-ahead to improve performance.

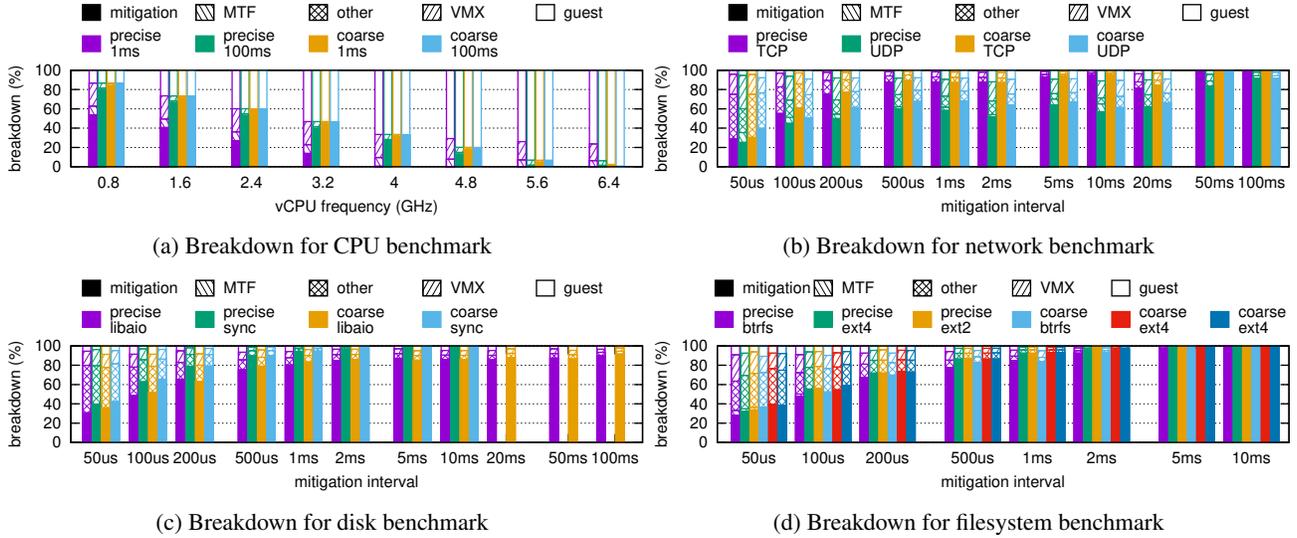

Figure 13: Execution time breakdowns, the whole execution time is 100%

*Filesystem* The filesystem benchmark also uses fio for three popular filesystems: btrfs, ext4 and ext2. We turn on LZO compression for btrfs, and journaling for both ext4 and btrfs. We simulates a common filesystem access pattern by performing random reads and writes on files. The file size ranges from 1KB to 1MB, and each I/O operation reads or writes between 4KB and 32KB of data. We use the `O_DIRECT` flag to skip the kernel data buffer, so as to reveal the impact of filesystems themselves.

Figure 11d and Figure 12d shows performance results and maximum leakage. Figure 13d breaks down execution time.

The ext filesystems show a pattern similar to the synchronous disk I/O benchmark, since most filesystem operations are synchronous. The journaling feature does not provide a performance benefit in this case.

The btrfs filesystem performs much better when the mitigation interval is long, Btrfs reads and writes the same amount of data with many fewer disk I/O operations, because of its compression feature and a better buffering mechanism. However, when the mitigation interval is short, its compression operations produce more overhead.

### 6.3 Macro Benchmarks

We use several benchmark suites to explore realistic cloud workloads. The PARSEC [11] benchmark represents scientific computation workloads. OLTP-Bench [17] represents typical relational database workloads. The YCSB [15] benchmark represents key-value storage workloads.

*PARSEC* Complementing the CPU micro-benchmark, PARSEC measures performance overhead for more complex compute-intensive applications. Instead of simple instructions, PARSEC applications exercise more CPU features such as FPU and cache. We compare Deterland's PARSEC performance against KVM with 3.2GHz real CPU speed.

The results are shown in Figure 14. The slowdown ratio is near 1 for all applications when vCPU speed is 6.4GHz. Most benchmarks show sub-linear slowdown ratios, as the actual functioning units are saturated, like FPU for `blackscholes` and `streamcluster`. The only application showing nearly linear slowdown is `fluidanimate`, which performs frequent lock operations; lock operations on a single-core vCPU degenerate to normal instructions. Performance is fairly competitive when the vCPU frequency is 3.2GHz, regardless of mitigation interval.

*Storage* We test the MySQL server with the TPC-C workload from OLTP-Bench, with local clients and remote clients. We also test the Cassandra server with three different workloads from YCSB, with local clients and remote clients. The backend filesystems for both MySQL and Cassandra are ext4 and btrfs, with the same configuration as in the filesystem micro benchmark. The remote clients run on a Linux machine outside the mitigation boundary.

Figure 15a shows performance overhead, and Figure 15b shows latency overhead. Overall performance is acceptable with a 1ms mitigation interval. Cassandra shows a better slowdown ratio than MySQL, as it has less chance of holding locks during mitigation. Cassandra is also more tolerant of mitigation level for local clients, as it has simpler buffering requirements than those of MySQL. Cassandra maintains the same slowdown ratio when the mitigation interval becomes 10ms, whereas MySQL becomes unusable.

The remote benchmark, however, reveals the impact of network latency. Both Cassandra and MySQL suffer significant performance drop when the mitigation level increases to 10ms. The average latency remains nearly the same for all local benchmarks, while the remote latency is higher due to extra network mitigation in addition to disk mitigation.

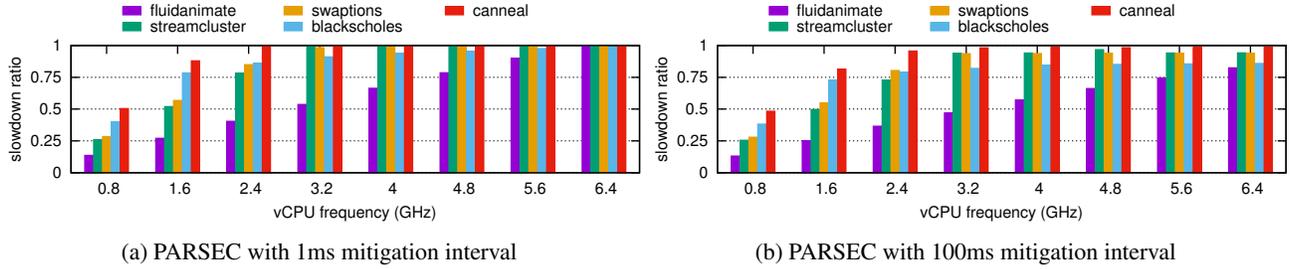

(a) PARSEC with 1ms mitigation interval

(b) PARSEC with 100ms mitigation interval

Figure 14: Performance overhead for scientific computation benchmarks

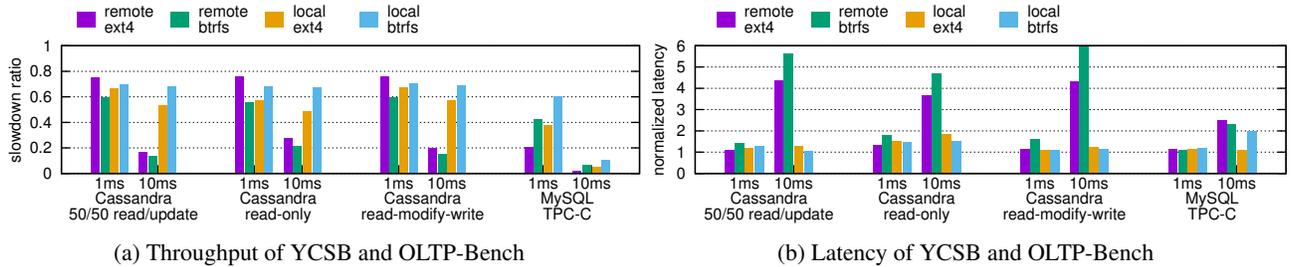

(a) Throughput of YCSB and OLTP-Bench

(b) Latency of YCSB and OLTP-Bench

Figure 15: Performance overhead for storage benchmarks

### 6.4 Deployment Considerations

As the benchmarks indicate, we find the best mitigation interval for legacy applications to be around 1ms. Estimated information leakage is around 100bps for all benchmarks, and theoretically bounded at 1Kbps. Of course, much of this estimated leakage is unlikely to be sensitive information useful to an attacker. If the VM sends network packets at high rate constantly, the current Deterland prototype spends more than 1ms to handle all the VM exits and data buffering, and exhibits a relatively fixed pattern of missed deadlines. This fixed pattern, if predictable, does not actually leak sensitive information. CPU-bounded applications exhibit little overhead. Network-bounded applications have no more than 50% throughput loss for both UDP and TCP, compared to KVM. Cloud providers such as Rackspace are known to limit network bandwidth for VM instances anyway, which may hide this throughput impact. While filesystem performance suffers, well-designed applications like databases handle this reasonably well and incur no more than 60% throughput loss in exchange for mitigation.

The ideal vCPU speed depends on the characteristics of the underlying physical CPU. However, most VM instances in the cloud do not execute a full allocation of instructions in a given segment. As a result, even if the vCPU speed is set high, the VM will likely be under-utilized.

In summary, applications that are CPU bounded may not notice any appreciable difference when running in Deterland. Applications using asynchronous I/O exhibit moderate overhead, less than 30%, with a 1ms mitigation interval. Applications that heavily depend on synchronous I/O may not be suitable for Deterland.

Deterland represents only one of many possible design points for systematic timing-channel mitigation in the cloud, of course, in many cases emphasizing generality and implementation simplicity over other considerations. It is likely that many further optimizations, possible hardware features, and specialization to particular application or security models of interest could further substantially improve Deterland's performance and practicality.

## 7. Related Work

Timing channel control has been well-studied in commercial security systems for decades [10, 12, 21, 28, 32, 39]. We now discuss related work grouped into three categories: controlling internal timing channels, controlling external channels, and detecting timing channels.

### 7.1 Controlling Internal Timing Channels

***Controlling cache-based timing channels*** Wang et al. propose an effort that refines the processor architecture to minimize information leakage through cache-based timing channels [49]. Oswald et al. present a method to disable cache sharing, avoiding cache-based timing attack [37]. these approaches are difficult to use in the cloud model, because either they require specific infrastructure architecture, or undermine the cloud's economic model of over-subscribing and statistically multiplexing resources among customers [6].

***System-enforced Determinism*** Aviram et al. propose to eliminate internal timing channels by forcing virtual machine execution to be deterministic [6]. The proposed approach does not address external timing channels, however. In addition, the effort achieves deterministic execution through Determinator, an OS that imposes new parallel programming constraints and limits compatibility with existing applications and OS kernels.

StopWatch [34] replicates each VM across three physical machines and uses the median of certain I/O events' timing from the three replicated VMs to determine the timings observed by other VMs co-resident on the same physical machines. However, StopWatch cannot fundamentally eliminate or bound internal timing channels, since attackers still can learn information by probing multiple victim VMs.

***Language-based timing channel control*** Many language-based efforts address internal timing channels [27, 52] and mitigate external timing channels [42, 44, 47]. Zhang et al. propose a language-level timing mitigation approach, which can reason about the secrecy of computation time more accurately [54]. In this effort, well-typed programs provably leak a bounded amount of information over time via external timing channels. Through predictive mitigation, this effort also offers an expressive programming model, where applications are expressible only if timing leakage is provably bounded by a desired amount. Unfortunately these language-based approaches require programs to be rewritten in new and unfamiliar languages.

### 7.2 Controlling External Timing Channels

***Additive noise*** To resist external timing channels, several proposals inject additive noise [23, 25, 26]. These approaches control timing channels only in specific workloads, however. In addition, stealthy timing channels robust to additive noise have been constructed [35].

***Predictive mitigation*** Predictive timing mitigation is a general scheme for provably limiting leakage through external channels [5, 53]. In general, predictive mitigation works by predicting future timing from past non-sensitive information, and enforcing these predictions. When a timing-sensitive event arrives before the predicted time point, the event is delayed to output according to the prediction; if a prediction fails, a new epoch with a longer prediction begins and information is leaked. This approach not only tracks the amount of information leaked at each epoch transition, but also offers a provable guarantee on total leakage. However, predictive mitigation does not directly address internal timing channels, a practical virtual machine's workload is hard to predict, and predictive mitigation cannot identify timing variations that may leak sensitive information, unnecessarily hurting performance when most timing variation is benign.

### 7.3 Detecting Timing Channels

Instead of controlling timing channels, some proposals aim to detect potential timing channels in programs or systems. Most existing approaches [13, 22, 38] detect timing channels either by inspecting some high-level statistic (*e.g.*, entropy) of network traffic, or by looking for specific patterns. A powerful adversary might be able to circumvent these defenses by varying timing in a slightly different way, however.

Chen et al. propose to use time-deterministic reply (TDR) to discover timing channels [14]. A TDR system not only offers deterministic replay, but also reproduces events that have non-deterministic timing, thus detecting potential timing channels. While both of these efforts make use of determinism, Deterland aims to control timing channels proactively rather than detecting timing channels retroactively.

## 8. Conclusion

This paper has presented Deterland, a hypervisor that runs unmodified OS kernels and applications while controlling both internal and external channels. While only a first step, our proof-of-concept prototype and experiments suggest that systematic timing channel mitigation may be practical and realistic for security-critical cloud applications, depending on configuration parameters and application workload.

## Acknowledgments

We thank our shepherd Ken Birman and the anonymous reviewers for their many helpful suggestions. We also thank Liang Gu and Daniel Jackowitz for their help in experiments, as well as David Wolinsky and Ennan Zhai for their help in copy-editing. This work was funded by NSF under grant CNS-1407454.

## References

[1] Onur Acıiçmez. Yet another microarchitectural attack: Exploiting I-cache. In *1st ACM Workshop on Computer Security Architecture (CCAW)*, November 2007.

[2] Onur Acıiçmez et al. Predicting secret keys via branch prediction. In *Cryptographers' Track - RSA Conference (CT-RSA)*, February 2007.

[3] Onur Acıiçmez, Werner Schindler, and Çetin K Koç. Cache based remote timing attack on the AES. In *Topics in Cryptology–CT-RSA 2007*, pages 271–286. Springer, 2006.

[4] Johan Agat. Transforming out timing leaks. In *Proceedings of the 27th ACM SIGPLAN-SIGACT Symposium on Principles of Programming Languages*, pages 40–53. ACM, 2000.

[5] Aslan Askarov, Danfeng Zhang, and Andrew C Myers. Predictive black-box mitigation of timing channels. In *Proceedings of the 17th ACM Conference on Computer and Communications Security*, pages 297–307. ACM, 2010.

[6] Amittai Aviram, Sen Hu, Bryan Ford, and Ramakrishna Gummadi. Determinating timing channels in compute clouds. In *Proceedings of the 2010 ACM Workshop on Cloud Computing Security Workshop*, pages 103–108. ACM, 2010.

[7] Amittai Aviram, Shu-Chun Weng, Sen Hu, and Bryan Ford. Efficient system-enforced deterministic parallelism. In *Proceedings of the 9th USENIX Conference on Operating Systems Design and Implementation*, pages 1–16. USENIX Association, 2010.

[8] Jens Axboe. Fio: Flexible i/o tester, 2008.

[9] James Matthew Barrie. *Peter and Wendy*. Hodder & Stoughton, 1911.

[10] Daniel J Bernstein. Cache-timing attacks on AES, 2005. http://cr.yp.to/antiforgery/cachetiming-20050414.pdf.

[11] Christian Bienia. *Benchmarking Modern Multiprocessors*. PhD thesis, Princeton University, January 2011.

[12] David Brumley and Dan Boneh. Remote timing attacks are practical. In *Proceedings of the 12th Conference on USENIX Security Symposium-Volume 12*. USENIX Association, 2003.

[13] Serdar Cabuk, Carla E Brodley, and Clay Shields. IP covert timing channels: Design and detection. In *Proceedings of the 11th ACM Conference on Computer and Communications Security*, pages 178–187. ACM, 2004.

[14] Ang Chen, W Brad Moore, Hanjun Xiao, Andreas Haeberlen, Linh Thi Xuan Phan, Micah Sherr, and Wenchao Zhou. Detecting covert timing channels with time-deterministic replay. In *USENIX Symposium on Operating System Design and Implementation (OSDI)*, 2014.

[15] Brian F Cooper, Adam Silberstein, Erwin Tam, Raghu Ramakrishnan, and Russell Sears. Benchmarking cloud serving systems with YCSB. In *Proceedings of the 1st ACM Symposium on Cloud Computing*, pages 143–154. ACM, 2010.

[16] Joseph Devietti, Brandon Lucia, Luis Ceze, and Mark Oskin. DMP: Deterministic shared memory multiprocessing. In *ACM SIGARCH Computer Architecture News*, volume 37, pages 85–96. ACM, 2009.

[17] Djellel Eddine Difallah, Andrew Pavlo, Carlo Curino, and Philippe Cudre-Mauroux. OLTP-Bench: An extensible testbed for benchmarking relational databases. *Proceedings of the VLDB Endowment*, 7(4), 2013.

[18] Yaozu Dong, Xiaowei Yang, Jianhui Li, Guangdeng Liao, Kun Tian, and Haibing Guan. High performance network virtualization with SR-IOV. *Journal of Parallel and Distributed Computing*, 72(11):1471–1480, 2012.

[19] George W. Dunlap et al. ReVirt: Enabling intrusion analysis through virtual-machine logging and replay. In *5th USENIX Symposium on Operating Systems Design and Implementation (OSDI)*, December 2002.

[20] George W Dunlap, Dominic G Lucchetti, Michael A Fetterman, and Peter M Chen. Execution replay of multiprocessor virtual machines. In *Proceedings of the 4th ACM SIGPLAN/SIGOPS International Conference on Virtual Execution Environments*, pages 121–130. ACM, 2008.

[21] Edward W Felten and Michael A Schneider. Timing attacks on web privacy. In *Proceedings of the 7th ACM conference on Computer and communications security*, pages 25–32. ACM, 2000.

[22] Steven Gianvecchio and Haining Wang. Detecting covert timing channels: an entropy-based approach. In *Proceedings of the 14th ACM Conference on Computer and Communications Security*, pages 307–316. ACM, 2007.

[23] J Giles and B Hajek. An information-theoretic and game-theoretic study of timing channels. *IEEE Transactions on Information Theory*, 48(9):2455–2477, 2002.

[24] Liang Gu, Alexander Vaynberg, Bryan Ford, Zhong Shao, and David Costanzo. CertiKOS: A certified kernel for secure cloud computing. In *Proceedings of the Second Asia-Pacific Workshop on Systems*, 2011.

[25] Andreas Haeberlen, Benjamin C. Pierce, and Arjun Narayan. Differential privacy under fire. In *20th Proceedings of USENIX Security Symposium (USENIX Security)*, August 2011.

[26] Wei-Ming Hu. Reducing timing channels with fuzzy time. In *IEEE Symposium on Security and Privacy*, pages 8–20, 1991.

[27] Marieke Huisman, Pratik Worah, and Kim Sunesen. A temporal logic characterisation of observational determinism. In *IEEE 19th computer Security Foundation Workshop (CSFW)*, July 2006.

[28] Ralf Hund, Carsten Willems, and Thorsten Holz. Practical timing side channel attacks against kernel space ASLR. In *IEEE Symposium on Security and Privacy*, pages 191–205. IEEE, 2013.

[29] Gerry Kane. *PA-RISC 2.0 Architecture*. Prentice Hall PTR, 1996.

[30] Richard A. Kemmerer. Shared resource matrix methodology: An approach to identifying storage and timing channels. *Transactions on Computer Systems (TOCS)*, 1(3):256–277, August 1983.

[31] Avi Kivity, Yaniv Kamay, Dor Laor, Uri Lublin, and Anthony Liguori. KVM: The Linux virtual machine monitor. In *Proceedings of the Linux Symposium*, volume 1, pages 225–230, 2007.

[32] Paul Kocher. Timing attacks on implementations of Diffie-Hellman, RSA, DSS, and other systems. In *Advances in Cryptology (CRYPTO)*, pages 104–113. Springer, 1996.

[33] Alexey Kopytov. Sysbench: A system performance benchmark, 2004.

[34] Peng Li, Debin Gao, and Michael K Reiter. StopWatch: a cloud architecture for timing channel mitigation. *ACM Transactions on Information and System Security (TISSEC)*, 17(2):8, 2014.

[35] Yali Liu, Dipak Ghosal, Frederik Armknecht, Ahmad-Reza Sadeghi, Steffen Schulz, and Stefan Katzenbeisser. Hide and seek in time – Robust covert timing channels. In *14th European Symposium on Research in Computer Security (ESORICS)*, September 2009.

[36] Gil Neiger, Amy Santoni, Felix Leung, Dion Rodgers, and Rich Uhlig. Intel virtualization technology: Hardware support for efficient processor virtualization. *Intel Technology Journal*, 10(3), 2006.

[37] Elisabeth Oswald, Stefan Mangard, Norbert Pramstaller, and Vincent Rijmen. A side-channel analysis resistant description

of the AES S-Box. In *Fast Software Encryption Workshop (FSE)*, February 2005.

[38] Pai Peng, Peng Ning, and Douglas S. Reeves. On the secrecy of timing-based active watermarking trace-back techniques. In *IEEE Symposium on Security and Privacy (S&P)*, May 2006.

[39] Colin Percival. Cache missing for fun and profit. In *BSDCan*, May 2005.

[40] Thomas Ristenpart, Eran Tromer, Hovav Shacham, and Stefan Savage. Hey, you, get off of my cloud: Exploring information leakage in third-party compute clouds. In *16th ACM Conference on Computer and Communications Security (CCS)*, pages 199–212, 2009.

[41] Rusty Russell. virtio: Towards a de-facto standard for virtual I/O devices. *ACM SIGOPS Operating Systems Review*, 42(5):95–103, 2008.

[42] Andrei Sabelfeld and David Sands. Probabilistic noninterference for multi-threaded programs. In *Computer Security Foundations Workshop*, pages 200–214, 2000.

[43] Claude Elwood Shannon and Warren Weaver. *The mathematical theory of communication*. University of Illinois Press, 1959.

[44] Geoffrey Smith and Dennis M. Volpano. Secure information flow in a multi-threaded imperative language. In *25th Symposium on Principles of Programming Languages (POPL)*, January 1998.

[45] Deian Stefan, Alejandro Russo, Pablo Buiras, Amit Levy, John C Mitchell, and David Maziéres. Addressing covert termination and timing channels in concurrent information flow systems. In *ACM SIGPLAN International Conference on Functional Programing (ICFP)*, 2012.

[46] Ajay Tirumala, Feng Qin, Jon Dugan, Jim Ferguson, and Kevin Gibbs. Iperf: Tcp and udp bandwidth measurement tool, 2005.

[47] Dennis M. Volpano and Geoffrey Smith. Eliminating covert flows with minimum typings. In *10th Computer Security Foundations Workshop (CSFW)*, June 1997.

[48] Zhenghong Wang and Ruby B. Lee. Covert and side channels due to processor architecture. In *22nd Annual Computer Security Applications Conference (ACSAC)*, December 2006.

[49] Zhenhong Wang and Ruby B. Lee. A novel cache architecture with enhanced performance and security. In *IEEE/ACM 41th International Symposium on Microarchitecture (Micro)*, 2008.

[50] Michael Weiss, Benedikt Heinz, and Frederic Stumpf. A cache timing attack on AES in virtualization environments. In *Financial Cryptography and Data Security*, pages 314–328. Springer, 2012.

[51] John C. Wray. An analysis of covert timing channels. In *IEEE Symposium on Security and Privacy*, May 1991.

[52] Steve Zdancewic and Andrew C Myers. Observational determinism for concurrent program security. In *Computer Security Foundations Workshop*, pages 29–43. IEEE, 2003.

[53] Danfeng Zhang, Aslan Askarov, and Andrew C Myers. Predictive mitigation of timing channels in interactive systems. In *Proceedings of the 18th ACM conference on Computer and communications security*, pages 563–574, 2011.

[54] Danfeng Zhang, Aslan Askarov, and Andrew C Myers. Language-based control and mitigation of timing channels. In *ACM SIGPLAN Conference on Programming Language Design and Implementation (PLDI)*, 2012.

[55] Xiao Zhang, Sandhya Dwarkadas, and Kai Shen. Towards practical page coloring-based multicore cache management. In *Proceedings of the 4th ACM European Conference on Computer Systems*, pages 89–102. ACM, 2009.

[56] Yinqian Zhang, Ari Juels, Alina Oprea, and Michael K Reiter. HomeAlone: Co-residency detection in the cloud via side-channel analysis. In *IEEE Security and Privacy (SP)*, pages 313–328, 2011.

[57] Yinqian Zhang, Ari Juels, Michael K Reiter, and Thomas Ristenpart. Cross-vm side channels and their use to extract private keys. In *Proceedings of the 2012 ACM Conference on Computer and Communications Security*, pages 305–316. ACM, 2012.